\documentclass[twocolumn,aip,reprint,superscriptaddress]{revtex4-2}

\usepackage{amsmath,amsfonts}
\usepackage{graphicx}% Include figure files

\usepackage{booktabs}
\usepackage{dcolumn}% Align table columns on decimal point

\usepackage{bm}% bold math
\usepackage{physics}
\usepackage{xcolor}
\usepackage{enumitem} 

\begin{document}

% \preprint{}

% \title{Spin-Lattice Relaxation under vibrational strong coupling: Comparing extend Dissipaton Equations of Motion simulations against Fermi's golden rule rate}
\title{Electronic frictional effects near metal surfaces with strong correlations}

\author{Yunhao Liu}
%\email[]{biruihao@westlake.edu.cn}
%\thanks{}
\affiliation{Department of Chemistry, School of Science and Research Center for Industries of the Future, Westlake University, Hangzhou, Zhejiang 310024, China}

\author{Wenjie Dou}%
\email{douwenjie@westlake.edu.cn} 
\affiliation{Department of Chemistry, School of Science and Research Center for Industries of the Future, Westlake University, Hangzhou, Zhejiang 310024, China}
\affiliation{Institute of Natural Sciences, Westlake Institute for Advanced Study, Hangzhou, Zhejiang 310024, China}
\affiliation{Key Laboratory for Quantum Materials of Zhejiang Province, Department of Physics, School of Science and Research Center for Industries of the Future, Westlake University, Hangzhou, Zhejiang 310024, China}

\date{\today}

\begin{abstract}
The electronic friction-Langevin dynamics (EF-LD) offers a simplified framework for describing nonadiabatic effects at metal surfaces, particularly in electrochemical and molecular electronic applications. We investigate the electronic friction behavior for the Hubbard-Holstein model using density matrix renormalization group (DMRG) theory. We show that electron-electron interactions lead to the formation of two energy levels in the impurity, resulting in two peaks in the electronic friction at the resonances of electron attachment or detachment with the metal's Fermi level. We further benchmark our results against mean field theory (MFT) and exact diagonalization (ED). The results calculated by ED and DMRG show strong agreement at high temperatures, suggesting the results from DMRG are reliable; however, at low temperatures, ED exhibits significant deviations relative to DMRG due to the finite-size limitations inherent in ED calculations. MFT completely fails to recover Fermi resonance in electronic friction. Moreover, we investigate the dynamics of the electronic friction using EF-LD. Simulations reveal differences between the electronic population and kinetic energy dynamics predicted by MFT and DMRG approaches, suggesting that  MFT approach is unreliable for nonadiabatic dynamics of strongly correlated systems.
\end{abstract}

\maketitle

%\tableofcontents
\section{\label{sec:sec1} Introduction}

When nuclear motion couples strongly to a continuum of electronic states (as encountered in molecule-metal surface interactions), the Born-Oppenheimer approximation fails strenuously. For example, see the experiments of gas molecules scattering off from a metal surface\cite{huang2000vibrational,doi:10.1126/science.aad4972}. While exact quantum mechanical treatments exist for closed systems with limited degrees of freedom, such as hierarchical equations of motion (HEOM)\cite{doi:10.1143/JPSJ.58.101,doi:10.1143/JPSJ.75.082001,10.1063/5.0011599}, multiconfiguration time-dependent Hartree (MCTDH)\cite{MEYER199073,10.1063/1.463007,Meyer2003}, such approaches prove computationally intractable for realistic interfacial systems. To overcome this limitation, the generalized Langevin dynamics framework has emerged as a practical computational paradigm for simulating nonadiabatic dynamics at the molecule-metal interfaces. In this framework, nonadiabatic dynamics is incorporated into electronic friction and random forces\cite{10.1063/1.440287,10.1063/1.432526}, thereby capturing essential electron transfer between the molecule and the metal surfaces while maintaining computational feasibility.

The electronic friction represents the first order correction to the Born-Oppenheimer approximation, providing a fundamental mechanism for understanding nonadiabaticity in a metallic bath. The concept of electronic friction has been widely applied in diverse fields, including electrochemistry \cite{doi:10.1021/jp805876e,doi:10.1021/acs.jpcc.5b06655,B805544K,doi:10.1021/jp9933673,PhysRevLett.84.1051} and molecular electronics\cite{doi:10.1021/jp7114548,Tao2006,Galperin_2007,PhysRevLett.100.176403,doi:10.1126/science.1081572,PhysRevB.88.201405,PhysRevB.76.085433}. For example, electronic friction has been used to explain the energy loss of gas molecules scattering off metal surfaces\cite{huang2000vibrational,doi:10.1021/acs.jpclett.5b02448,C0CP02086A}. Moreover, electronic friction has also been shown to be relevant for chemisorption and dissociation\cite{doi:10.1126/science.aad4972,PhysRevLett.118.256001,PhysRevLett.117.196001}. That being said, the electronic friction is mostly calculated for non-interacting electrons. The inclusion of electron-electron is challenging \cite{PhysRevB.88.045137,https://doi.org/10.1002/smll.200600101,PhysRevLett.94.206804,Kisiel2011,Langer2014,PhysRevB.60.5969,Kennes2017}.

Let us consider the coupled electron-nuclear motion with el-el interaction. The total Hamiltonian $\hat{H}_\text{tot}$ can be divided into the electronic Hamiltonian $\hat{H}\left(\mathbf{R}\right)$, and nuclear kinetic energy
\begin{equation}\label{eqn-1}
\hat{H}_{\mathrm{tot}}=\hat{H}\left(\mathbf{R}\right)+\sum_{\alpha} \frac{P_{\alpha}^{2}}{2 m_{\alpha}}+ U_0\left(\mathbf{R}\right),
\end{equation}
where $U_0\left(\mathbf{R}\right)$ represents the classical nuclear repulsion potential, and the electronic Hamiltonian $\hat{H}\left(\mathbf{R}\right)$ consists of a manifold of electron,
\begin{equation}\label{eqn-2}
\hat{H}\left(\mathbf{R}\right)=\sum_{pq}h_{pq}\hat{c}_p^\dagger\hat{c}_q + \sum_{pqrs}g_{pqrs}\hat{c}_p^\dagger\hat{c}_q^\dagger\hat{c}_r\hat{c}_s.
\end{equation}
Theoretical frameworks for electronic friction originate from considering nuclei interacting with fast-relaxing electronic baths. In this paradigm, nuclear motion perturbs the electronic equilibrium. Due to the fast relaxation of the electronic bath compared to nuclear motion, the resulting feedback force on the nuclei, while fundamentally arising from a delayed electronic response, can be modeled effectively through Markovian damping and thermal noise\cite{PhysRevLett.119.046001}. The resulting Langevin equation governing nuclear coordinates ($\mathbf{R}$) takes the form:
\begin{equation}\label{eqn-3}
m_{\alpha} \ddot{R}_{\alpha}=\bar{F}_{\alpha}-\sum_{v} \gamma_{\alpha v} \dot{R}_{v}+\zeta_{\alpha}(t),
\end{equation}
where $\alpha$ and $\nu$ are index nuclear degrees of freedom (DoFs), $\bar{F}_\alpha = -\operatorname{tr}_e\left(\partial_\alpha\hat{H}\rho_\mathrm{ss}\right)$ represents the mean force, $\gamma_{\alpha\nu}$ the (Markovian) friction tensor, and $\zeta_{\alpha}(t)$ the random force component. In our previous work, we have derived a universal electronic friction from a quantum-classical Liouville equation (QCLE)\cite{PhysRevLett.119.046001}, which should be valid in and out of the electronic equilibrium, with or without el-el interactions:
\begin{align}\label{eqn-4}
\gamma_{\mu \nu}(\mathbf{R})=-\int_{0}^{\infty} \mathrm{d} t \operatorname{tr}_{e}&\left[\partial_{\mu} \hat{H}(\mathbf{R}) e^{-\mathrm{i} \hat{H}(\mathbf{R}) t / \hbar}\right.   \nonumber \\
&\left.\times\partial_{\nu} \hat{\rho}_{s s}(\mathbf{R}) e^{\mathrm{i} \hat{H}(\mathbf{R}) t / \hbar}\right],
\end{align}
where $\mu$ and $\nu$ are nuclear DoFs, $\hat{H}\left(\mathbf{R}\right)$ is the electronic Hamiltonian, $\hat{\rho}_{ss}\left(\mathbf{R}\right)$ is the steady state's electronic density matrix, and $\mathrm{tr}_e$ implies tracing over many-body electronic states. Subsequently, we have derived an explicit, very general formula for calculating that friction tensor in and out of the electronic equilibrium\cite{Bode2012,PhysRevB.96.104305,PhysRevB.97.064303} under the premise that the el-el interactions are independent of position in Eq. (\ref{eqn-2}):
\begin{equation}\label{eqn-5}
\gamma_{\mu \nu}=\hbar \int \frac{\mathrm{d} \epsilon}{2 \pi} \operatorname{tr}_{s}\left(\partial_{\mu} \mathcal{H}\left(\mathbf{R}\right) \partial_{\epsilon} G^{R} \partial_{\nu} \mathcal{H}\left(\mathbf{R}\right) G^{<}\right)+\text {H.c.},
\end{equation}
where $\mathrm{tr}_s$ implies summation over system orbitals, $G^R$ and $G^<$ are system retard Green's functions (GF) and lesser GFs respectively, and $\mathcal{H}\left(\mathbf{R}\right)\equiv \sum_{pq}h_{pq}\hat{c}_p^\dagger\hat{c}_q$ is the quadratic Hamiltonian in Eq. (\ref{eqn-2}). In time domain, system retarded GFs and lesser GFs are defined as 
\begin{equation}\label{eqn-6}
G^<_{qp}\left(t_1,t_2\right)=\frac{\mathrm{i}}{\hbar}\left\langle \hat{c}_p^\dagger\left(t_2\right)\hat{c}_q\left(t_1\right)\right\rangle_{\mathbf{R}\left(t_1\right)},
\end{equation}
and
\begin{equation}\label{eqn-7}
G^R_{qp}\left(t_1,t_2\right)=-\frac{\mathrm{i}}{\hbar}\theta\left(t_1-t_2\right)\left\langle\left\{\hat{c}_q\left(t_1\right),\hat{c}_p^\dagger\left(t_2\right)\right\}\right\rangle_{\mathbf{R}\left(t_1\right)}
\end{equation}
respectively. 

At the electronic equilibrium, Eq. (\ref{eqn-5}) can be simplified as\cite{PhysRevB.96.104305}
\begin{equation}\label{eqn-8}
\gamma_{\mu \nu}=-\pi \hbar \int \mathrm{d} \epsilon \operatorname{tr}_s\left(\partial_{\mu} \mathcal{H}\left(\mathbf{R}\right) P \partial_{\nu} \mathcal{H}\left(\mathbf{R}\right) P\right) \partial_{\epsilon} f,
\end{equation}
where $f$ is the Fermi function, $f\left(\epsilon\right)=\left(\mathrm{e}^{\beta\epsilon}+1\right)^{-1}$, $P\equiv -\frac{1}{\pi}\operatorname{Im}G^{R}$.

In previous work, the electronic friction with el-el interactions have been evaluated, which can give very new and interesting physics at low temperature\cite{PhysRevB.58.2191,PhysRevB.60.5969,PhysRevLett.119.046001}. In Ref. \cite{PhysRevLett.119.046001}, the Anderson model is calculated by numerical renormalization group (NRG)\cite{RevModPhys.80.395}. However, the bath is still assumed to be non-interacting. In this work, we evaluate the electronic friction of interacting electrons from the Hubbard model with density matrix renormalization group (DMRG)\cite{PhysRevLett.69.2863,PhysRevB.48.10345,RevModPhys.77.259,SCHOLLWOCK201196}. We compare our results to mean field theory (MFT), exact diagonalization (ED). We show that the DMRG is reliable in predicting electronic friction whereas MFT can give rise to wrong dynamics. 

The structure of this paper is as follows. In Sec.\ref{sec:sec2}, we give a brief introduction to the framework of DMRG to evaluate correlation function in time domain and frequency domain, and a general model for the study of strongly correlated systems with electron-phonon (el-ph) coupling. In Sec.\ref{sec:sec3-1}, we compare the electronic friction according to different methods, and we compare the results where the mean force and electronic friction evaluated by MFT and DMRG; in Sec.\ref{sec:sec3-2}, we study the dynamics of the electronic friction. We conclude in Sec.\ref{sec:sec4}.

\section{\label{sec:sec2} Methodology}

The Hubbard-Holstein (HH) model provides an ideal testbed for studying strongly correlated systems with electron-phonon (el-ph) coupling. To establish the importance of el-el interactions for every system and bath site, we will now calculate the electronic friction for the HH model,
\begin{widetext}
\begin{align}
\hat{H}&=\hat{H}_\mathrm{Hub}+\hat{H}_\mathrm{osc},  \label{eqn-9} \\
\hat{H}_\mathrm{Hub}&=E(x)\sum_\sigma \hat{c}_{1\sigma}^\dagger \hat{c}_{1\sigma}+\epsilon \sum_{i\ne 1,\sigma}\hat{c}_{i\sigma}^\dagger \hat{c}_{i\sigma}+t\sum_{i,\sigma}\left(\hat{c}_{i\sigma}^\dagger \hat{c}_{i+1,\sigma} + \text{h.c.}\right) + U\sum_i\hat{n}_{i\uparrow}\hat{n}_{i\downarrow},   \label{eqn-10} \\
\hat{H}_\mathrm{osc}&= \frac{p^2}{2m} + \frac{1}{2}\omega^2x^2,   \label{eqn-11}
\end{align}
\end{widetext}
where $\hat{c}^\dagger_{i\sigma}$,$\hat{c}_{i\sigma}$ are fermionic creation and annihilation operators of spin $\sigma$ on site $i$, and $\hat{n}_{i\sigma}\equiv\hat{c}_{i\sigma}^\dagger\hat{c}_{i\sigma}$. Physically, the Hubbard-Holstein model represents an electronic impurity on 1-th site near a bath and coupled to a vibrating oscillator with position and momentum $x$ and $p$. The impurity can be filled with an electron of up or down spin, such that  $\sigma = \uparrow,\downarrow$ indicates spin states. The oscillator is a vibrational DoF and feels a different force depending on the occupation of the impurity. We set the on-site energy of the impurity in Eq. (\ref{eqn-10}) to be $E(x)\equiv E_d + \sqrt{2}gx$. To understand how the motion of the oscillator is perturbed by the fluctuating charge of the impurity, we will evaluate the electronic friction of Hubbard-Holstein model at different temperatures.

We use Eq. (\ref{eqn-8}) to calculate the electronic friction tensor in our DMRG calculation, since the system has one electronic bath instead of two electronic baths with different temperatures or chemical potentials.

\subsection{Time-dependent DMRG for finite-temperature calculations}
The computation of retarded Green's functions (Eq. (\ref{eqn-7})) at finite temperatures employs time-dependent DMRG\cite{PAECKEL2019167998} (TD-DMRG) techniques combining imaginary- and real-time propagation. Our implementation utilizes the purification approach \cite{PhysRevLett.93.207204} for thermal state representation, where mixed states are encoded as pure states in an enlarged Hilbert space P$\otimes$Q formed by adding an auxiliary space Q to the physical space P. Thermal equilibrium states at specific temperature $\left|\psi_\beta\right\rangle$ are generated through imaginary-time evolution from the maximally entangled identity state:
\begin{align}
\left|\psi_\beta\right\rangle&=\frac{1}{\sqrt{Z}}\mathrm{e}^{-\beta\hat{H}}\left|I\right\rangle,  \label{eqn-12} \\
\left|I\right\rangle&\equiv\sum_n\left|n,\tilde{n}\right\rangle,  \label{eqn-13}
\end{align}
where $\left|\tilde{n}\right\rangle$ is the state in auxiliary space that is same as the state $\left|n\right\rangle$ in physical space.

For time evolution operators, we adopt a hybrid strategy leveraging both time-evolving block decimation (TEBD) \cite{PhysRevLett.91.147902,PhysRevLett.93.040502} and time-dependent variational principle (TDVP) \cite{PhysRevLett.107.070601,PhysRevB.94.165116,PAECKEL2019167998} methods. For any nearest-neighbor Hamiltonian with $2N$ sites
\begin{equation}
    \hat{H}=\sum_{j=1}^{2N-1}\hat{h}_{j,j+1},
\end{equation}
where $\hat{h}_{j,j+1}$ acts on the $j$-th site and $(j+1)$-th site. The Hamiltonian can be decomposed into two parts 
\begin{align}
    \hat{H} &= \hat{H}_1 + \hat{H}_2,  \\
    \hat{H}_1 &= \sum_{j=1}^{N}\hat{h}_{2j-1,2j},  \\
    \hat{H}_2 &= \sum_{j=1}^{N-1}\hat{h}_{2j,2j+1}.
\end{align}
The TEBD algorithm employs a second-order Suzuki-Trotter decomposition:
\begin{equation}\label{eqn-18}
\mathrm{e}^{-\tau\hat{H}}=\mathrm{e}^{-\tau\hat{H}_{1}/2} \mathrm{e}^{-\tau\hat{H}_{2}} \mathrm{e}^{-\tau\hat{H}_{1}/2}+\mathcal{O}\left(\tau^{3}\right).  
\end{equation}
While TEBD efficiently handles nearest-neighbor interactions in Hubbard model, its long-time accuracy suffers from non-unitary errors. To address this limitation, we implement TDVP-based propagation governed by the variational condition:
\begin{equation}\label{eqn-19}
\min \left \| H\left|\psi(t)\right\rangle -\mathrm{i}\hbar\frac{\partial}{\partial t} \left|\psi(t)\right\rangle \right \|,
\end{equation}
which preserves wavefunction norms through constrained optimization in the matrix product state (MPS) manifold \cite{PAECKEL2019167998}.

The TDVP framework offers two implementation schemes: (i) two-site formulation enabling dynamic bond dimension adaptation during temporal evolution, and (ii) single-site approach maintaining fixed bond dimensions for computational efficiency. These variants exhibit contrasting resource scaling: $\mathcal{O}\left(Nm^3dt\right)$ for single-site versus $\mathcal{O}\left(Nm^3d^2t\right)$ for two-site implementations, where $N$ represents lattice sites, $m$ the virtual bond dimension, and $d$ the physical bond dimension. For spinful electronic systems, $d=4$.

\subsection{Dynamical DMRG for zero-temperature calculations}
The temperature dependence of correlation effects introduces distinct computational challenges: enhanced electronic correlations at low temperatures prolong the relaxation timescale of retarded Green's functions (GFs), necessitating extended temporal propagation for accurate characterization. This requirement conflicts with the inherent difficulty in simulating long-time dynamics due to entanglement entropy's quasi-linear temporal growth \cite{Calabrese_2005}—a phenomenon demanding exponentially increasing bond dimensions for error control. To circumvent this dichotomy, frequency-domain approaches like dynamical DMRG (DDMRG) \cite{PhysRevB.66.045114} provide an alternative pathway by directly computing spectral properties.

At zero temperature, the Fermi function derivative collapses to a delta distribution:
\begin{equation}\label{eqn-20}
\lim_{\beta \to \infty} \partial_\epsilon f = -\delta\left(\epsilon\right),
\end{equation}
so we can only calculate the retarded GFs when $\epsilon=0$ in frequency domain at zero temperature to evaluate electronic friction (Eq. (\ref{eqn-8})) according to dynamical DMRG (DDMRG).

For a general correlation function $C\left(\omega\right)=\int\left \langle A^\dagger(t)A \right \rangle \mathrm{e}^{\mathrm{i}\omega t}\mathrm{d}\omega$, in general, we are usually interested in calculating the imaginary part of the correlation function
\begin{align}\label{eqn-21}
I(\omega) &\equiv \mathrm{Im}C(\omega)  \nonumber \\
&= -\left\langle\psi_0 \right| A^\dagger\frac{\eta}{\left(\hbar\omega-H+E_0\right)^2+\eta^2}A\left | \psi_0  \right \rangle , 
\end{align}
where $H$ is the time-independent Hamiltonian, $E_0$ and $\left|\psi_0\right\rangle$ are ground-state energy and wavefunction, $A$ is the operator corresponding to the physical quantity which is analyzed. A small real number $\eta >0$ is used in the calculation to shift the poles of the correlation function into the complex plane. The imaginary part of correction vector associated with $C(\omega)$ is defined by
\begin{align}\label{eqn-22}
\left|X(\omega)\right\rangle &\equiv \mathrm{Im}\left\{\frac{1}{\hbar\omega-H+E_0+\mathrm{i}\eta}A\left|\psi_0\right\rangle\right\}  \nonumber \\  
&= -\frac{\eta}{\left(\hbar\omega-H+E_0\right)^2+\eta^2}A\left | \psi_0  \right \rangle.
\end{align}
The computational kernel involves solving the inhomogeneous equation:
\begin{equation}\label{eqn-23}
\left[\left(\hbar\omega-H+E_0\right)^2+\eta^2\right]\left|X(\omega)\right\rangle=-\eta A\left|\psi_0\right\rangle.
\end{equation}
where the correction vector $X(\omega)$ encapsulates the dynamical response. This is equivalent to minimizing the constrained functional:
\begin{widetext}
\begin{align}\label{eqn-24}
\mathcal{F}\left(\omega\right) = \left\langle X(\omega)\right| \left[\left(\hbar\omega-H+E_0\right)^2+\eta^2\right]\left|X(\omega)\right\rangle + \eta\left\langle\psi_0\right|A^\dagger\left|X(\omega)\right\rangle + \eta \left\langle X(\omega)\right|A\left|\psi_0\right\rangle,
\end{align}
\end{widetext}
establishing a variational framework for spectral resolution across frequency space.

We summarize our strategy for evaluating the retarded Green's functions (Eq.~(\ref{eqn-7})) according to DMRG as follows:

\textbf{Finite temperature:}
\begin{enumerate}[noitemsep, topsep=0pt, partopsep=0pt, parsep=0pt, itemindent=*, leftmargin=*]
    \item Obtain the thermal equilibrium state \(\left|\psi_\beta\right\rangle\) via imaginary-time evolution starting from the maximally entangled identity state (Eq.~(\ref{eqn-12})).
    \item Evaluate the Green's function through real-time evolution.
\end{enumerate}
During both evolution stages, the evolution workflow is as follows\cite{PhysRevX.11.031007}:
\begin{enumerate}[noitemsep, topsep=0pt, partopsep=0pt, parsep=0pt, itemindent=*, leftmargin=*]
    \item Apply the TEBD with high accuracy up to a time $\tau_\text{TEBD}$ to obtain a MPS with a suitably large bond dimension.
    \item Employ two-site TDVP (2TDVP) to further increase the bond dimension.
    \item Upon reaching maximum bond dimension, switch to single-site TDVP (1TDVP) for computational efficiency.
\end{enumerate}

\textbf{Zero temperature:}
\begin{enumerate}[noitemsep, topsep=0pt, partopsep=0pt, parsep=0pt, itemindent=*, leftmargin=*]
    \item Calculate the ground-state wavefunction by DMRG.
    \item Evaluate the Green's function using DDMRG in the frequency domain.
\end{enumerate}  

All DMRG calculations were performed using the ITensor library\cite{itensor}, leveraging its native implementations of TEBD and TDVP algorithms. To enable finite-temperature simulations, we extended the ITensor framework through purification method. We developed a DDMRG module building upon ITensor's tensor contraction engine and matrix product state (MPS) infrastructure to implement our calculations.

\section{\label{sec:sec3} Results and Discussions}

\subsection{Electronic friction and potential of mean force}\label{sec:sec3-1}

\begin{figure*}[htbp]
    \centering
    \includegraphics[width=0.75\linewidth]{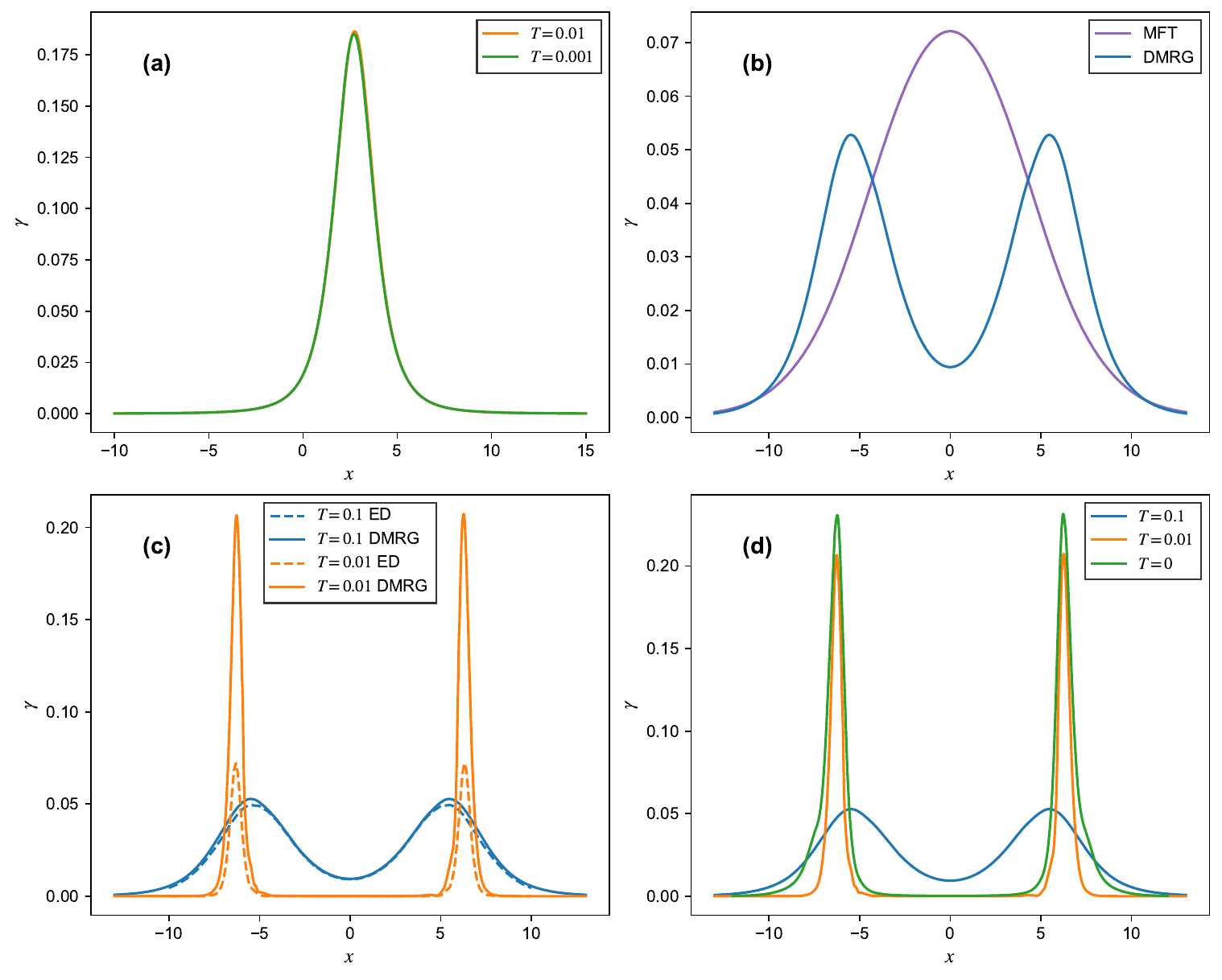}
    \caption{\label{fig-1}{(a) Electronic friction of the lattice model without el-el interactions ($U=0$) as a function of position $x$. Only one peak occurs in the electronic friction. (b) Electronic friction according to MFT and DMRG calculations at temperature $T=0.1$. Since there are el-el repulsions in the Hubbard model, DMRG successfully predicts two peaks in the friction, while MFT fails to recover the two peaks. (c) Electronic friction according to DMRG and ED. When $T=0.1$, the results of 10-site Hubbard model calculated by ED are basically consistent with DMRG, and the finite-size effects are not yet very obvious at this temperature; when the temperature $T=0.01$, there is a significant difference between the results of ED and DMRG, and the finite-size effects have a great impact on the results. (d) Electronic friction as a function of position $x$ according to DMRG. In all of our calculations above, the parameters are $E_d = -0.5$, $\epsilon=-0.5$, $g = 0.075$, $t = 0.3$, $U = 1$, and we set $k_B = \hbar = 1$.}}
\end{figure*}

In Fig. \ref{fig-1}(a), we show that the electronic friction as a function of $x$ of Hubbard-Holstein model (Eqs. (\ref{eqn-9})-(\ref{eqn-11})) without el-el interactions ($U=0$). There is one peak in the electronic friction, where there is the resonance of electron attachment or detachment with the Fermi level of the metal $\epsilon_F$, i.e. $E_d + \sqrt{2}gx=0$ (we have set $\epsilon_F=0$). Notice that, as we decrease temperature $T=0.01$ to $T=0.001$, the friction does not change, meaning we have reached to the zero temperature limit. In Fig. \ref{fig-1}(b), we compare the electronic friction calculated from DMRG versus the results from MFT at temperature $T=0.1$. Notice that DMRG predicts two peaks in the electronic friction, indicating the existence of a new energy level due to el-el interactions. Therefore, resonances of electron attachment or detachment in impurity with the Fermi level of other sites within the Hubbard model occur near $E_d + \sqrt{2}gx=0$ and $E_d + \sqrt{2}gx+U=0$. In contrast, MFT only predicts one peak at the position in the middle of two Fermi resonance, where DMRG predicts a dip. In Fig. \ref{fig-1}(c), we compare the electronic friction calculated from DMRG versus the results calculated from ED at different temperatures. We have limited the calculation to 10 sites in ED due to the high computational cost, while 20-site Hubbard model is calculated in DMRG. At $T=0.1$, the results of Hubbard model calculated by ED agree with DMRG very well; while at lower temperature $T=0.01$, the difference between ED and DMRG is larger. When simulating lattice models like the Hubbard model, approaching the thermodynamic limit (infinite system size) is crucial for obtaining physically meaningful results that represent bulk materials. The required number of lattice sites $L$ to achieve this approximation depends significantly on temperature. Finite-size effects refer to deviations in the properties of a simulated system from those of the thermodynamic limit due to the system's finite size. Generally, the finite-size effects are more significant at low temperatures than at high temperatures, so larger system sizes are required at low temperatures compared to high temperatures. the results in Fig. \ref{fig-1}(c) are consistent with our expectations. Approaches such as DMRG to calculate large systems are necessary for low temperature. In Fig. \ref{fig-1}(d), we have further included zero temperature results for electronic friction from DMRG. The electronic frictions at temperature $T=0.1$ and $T=0.01$ are calculated by TD-DMRG, and the results at temperature $T=0$ are calculated by DDMRG. As mentioned above, at lower temperatures the correlation effects are stronger, such that the time for the real-time Green's functions to decay to 0 is very long. DDMRG is more suitable to evaluate the electronic friction at zero temperature. Notice that the results from TD-DMRG is consistent with DDMRG at low temperature. 

In the diabatic picture, there are three different potential energy surfaces (PESs)-those with the impurity unoccupied (denoted as 0), those with the impurity occupied by only one electron (denoted as 1), and those with the impurity occupied by two electrons (denoted as 2):
\begin{align}
H_{\alpha} & =\frac{p^2}{2m} + V_{\alpha}, \alpha=0,1,2 \\
V_{0} & =\frac{1}{2} \omega^2 x^{2}, \\
V_{1} & =\frac{1}{2} \omega^2 x^{2} + E(x), \\
V_{2} & =\frac{1}{2} \omega^2 x^{2} + 2E(x) + U.
\end{align}
Also, the potential of mean force (PMF) is defined as\cite{Bode2012}
\begin{equation}\label{eqn-29}
V_\text{PMF}=\frac{1}{2}\omega^2 x^2 - \int_{x_0}^x\mathrm{d}x^\prime \bar{F}_x\left(x^\prime\right),
\end{equation}
where $\bar{F}_x$ is the mean force along $x$ direction defined as Eq. (\ref{eqn-3}).

\begin{figure*}[htbp]
    \centering
    \includegraphics[width=0.75\linewidth]{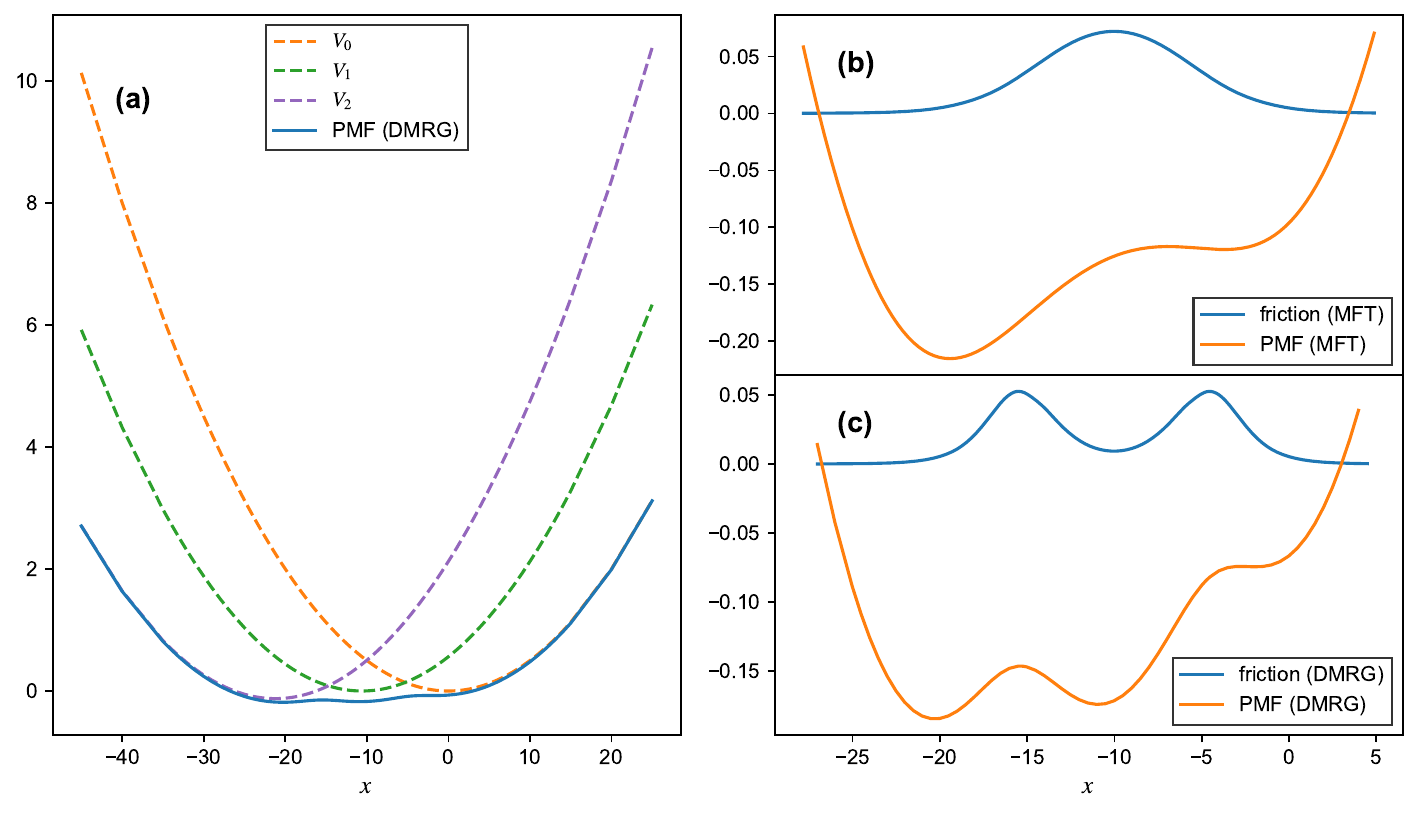}
    \caption{\label{fig-2}{(a) PMF and diabatic PESs as a function of position. (b) PMF and electronic friction as a function of position according to MFT. (c) PMF and electronic friction as a function of position according to DMRG.  The friction shows two peaks where the PMF reaches its maximum value where the PESs $V_0$ and $V_1$, as well as $V_1$ and $V_2$ cross according to DMRG. The parameters are $g = 0.075$, $\omega = 0.1$, $t = 0.3$, $U = 1$, $E_d=\frac{g^2}{m\omega^2}$, $T=0.1$, and we set $m = k_B = \hbar = 1$.}}
\end{figure*}

In Fig. \ref{fig-2}(a), we plot the potential of mean force calculated by DMRG and potential energy surfaces as a function of $x$ for the HH model. At the intersection of potential energy surfaces $V_0$ and $V_1$, as well as $V_1$ and $V_2$, the potential of mean force shows two peaks. Again, the peaks indicate the Fermi resonance. In Fig. \ref{fig-2}(b) and (c), we plot the potential of mean force and electronic friction as a function of $x$ calculated by MFT and DMRG respectively. Note that the PMF shows two peaks in the same position with the peaks shown in the electronic frictions. At these points, electrons are exchanged between the molecule and metal surface most frequencly, where the partial occupation of the molecule facilitates this exchange. But MFT only predicts one peak in the electronic friction as well as in the potential of mean force. This result indicates that the MFT fails in predicting correct PMF and electronic friction for strongly correlated systems. Here we set $E_d=\frac{g^2}{m\omega^2}$ to conveniently demonstrate the relationship between PESs, PMF and friction.
 
\subsection{Electronic friction-Langevin dynamics(EF-LD)}\label{sec:sec3-2}

We will now study the nonadiabatic dynamics within the electronic frictional model. In our previous work\cite{10.1063/1.4927237} comparing the dynamics according to surface hopping (SH) and EF-LD, we observed significant disagreement between the two approaches in the short term, with agreement emerging only in the long term. EF-LD proved significantly more reliable for nuclear observables than the impurity population.

The Langevin equation (\ref{eqn-3}) can be simplified as follows in the one-dimensional case,
\begin{align}
m \dot{v}_x&=\bar{F}_{x} -\gamma{v}_{x}+\zeta(t),  \label{eqn-30} \\
v_x&=\frac{\mathrm{d}x}{\mathrm{d}t},   \label{eqn-31} 
\end{align}
where the random force $\zeta\left(t\right)$ is assumed to be a Gaussian variable with a norm $\sigma = \sqrt{\frac{2\gamma mkT}{\mathrm{d} t}}$. This condition satisfies the second fluctuation-dissipation theorem\cite{10.1063/1.4733675}. $\mathrm{d} t$ is the time step interval. We then use 4th order Runge-Kutta (RK4) to integrate Eqs. (\ref{eqn-30}) and (\ref{eqn-31}), where 10000 trajectories have been used for the EF-LD simulations. We initialize the oscillators localized in one of the wells at $x=0$ by sampling their states from a Boltzmann distribution, where the average initial kinetic energy per oscillator is $5k_\mathrm{B}T$. The random force $\zeta(t)$ is generated by a normal distribution.    

\begin{figure*}[htbp]
    \centering
    \includegraphics[width=0.75\linewidth]{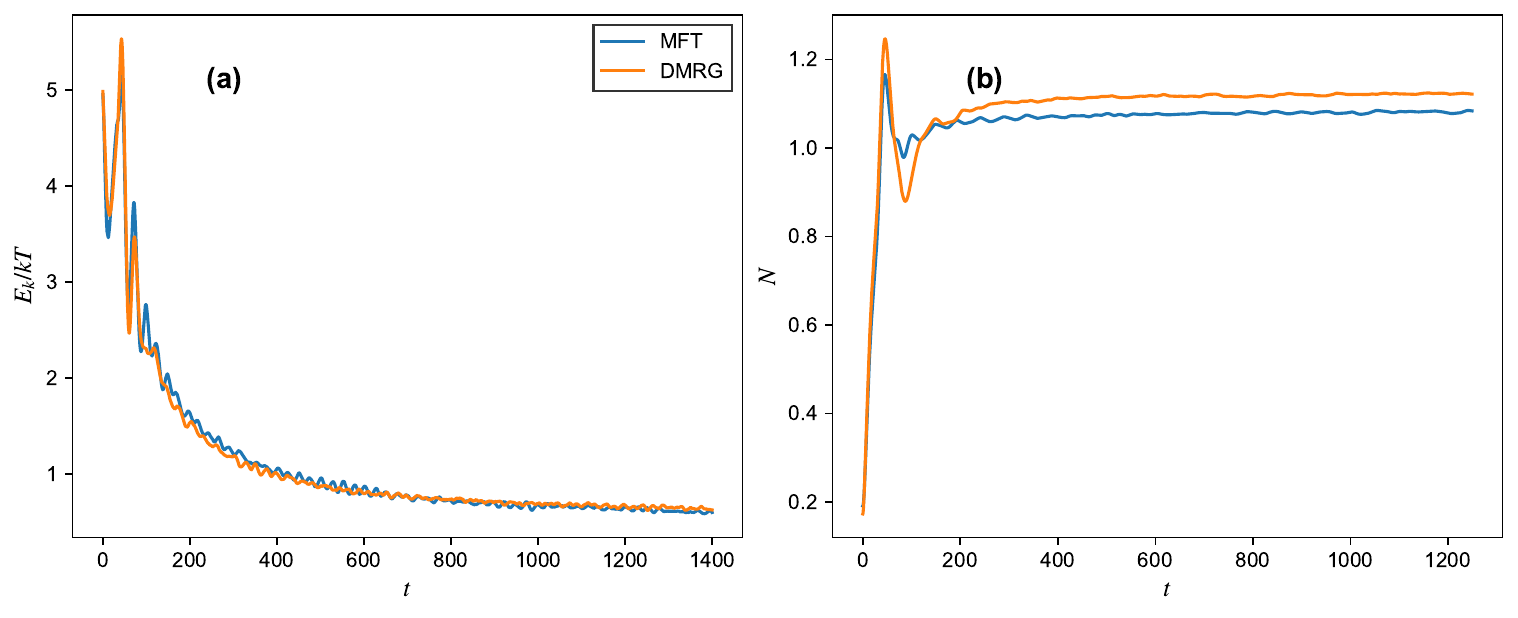}
    \caption{\label{fig-3}{The results from EF-LD. (a) average kinetic energy; (b) electronic population in the impurity. The parameters are $g = 0.075$, $\omega = 0.1$, $t = 0.3$, $U = 1$, $E_d=\frac{g^2}{m\omega^2}$, $T=0.1$, and we set $m = k_B = \hbar = 1$. We initialize the oscillators localized in one of the wells at $x=0$ by sampling their states from a Boltzmann distribution, where the average initial kinetic energy per oscillator is $5k_\mathrm{B}T$.}}
\end{figure*}

Figure \ref{fig-3} presents the EF-LD results for both the average kinetic energy and the electronic population of the impurity. We carry out the EF-LD simulation using the electronic friction and potential of mean force from both MFT and DMRG calculations. According to the equipartition theorem, the long-time average kinetic energy should converge to $\frac{1}{2}kT$. Both MFT- and DMRG-based results eventually approach this limit. Notice also that the dynamics for nuclear motion from  MFT and DMRG are in a good agreement. However, substantial discrepancies arise between the two methods for electronic dynamics-at short times, the DMRG-based electronic population is different from the MFT-based; at longer time, the equilibrium population also exhibits significantly differences between the two methods. These disagreements can be understood by examining Fig. \ref{fig-2}, which shows significant differences between the MFT- and DMRG-derived mean forces and electronic friction coefficients in the Langevin equation (Eq. (\ref{eqn-3}) or (\ref{eqn-30})). Consequently, we expect that the dynamical results from DMRG and MFT are very different.

\section{\label{sec:sec4} Conclusion}

In this work, we systematically evaluate electronic friction in the Hubbard-Holstein model as a function of the impurity position $x$, employing mean field theory (MFT), exact diagonalization (ED) and density matrix renormalization group (DMRG). The electronic friction exhibits two distinct peaks associated with electron attachment and detachment resonances at the metal's Fermi level due to electron-electron interactions. While MFT fails to capture these two peaks, highlighting its limitations for strongly correlated systems, both ED and DMRG accurately reproduce the two peaks.

As temperature decreases, the magnitudes of the peaks increase, and finite-size effects become more pronounced, necessitating large-scale methods like DMRG for accurate low-temperature calculations. Physically, the peaks in electronic friction occur when electron-exchange processes are most active near the Fermi energy, where the partial occupation of metal electron levels enhances such transitions. 

Although the MFT-based and DMRG-based electronic friction-Langevin dynamics (EF-LD) of average kinetic energy agree well with each other, the inaccuracies in MFT-based electronic friction and mean force arise in the dynamics of electronic population: MFT-based EF-LD deviates substantially from the more reliable DMRG-based results. This proves the importance of beyond MFT approaches for modeling dynamics of strongly correlated systems. Notice that, for the case without el-el repulsion, the lattice model is equivalent to the impurity model. The EF-LD method for the impurity model has been benchmarked in Refs. \cite{10.1063/5.0204307}. Moreover, the cases with el-el repulsion from DMRG-based EF-LD should be validated against numerically exact results as well.

\begin{acknowledgments}
W.D. thanks the funding from National Natural Science Foundation of China (No. 22361142829) and Zhejiang Provincial Natural Science Foundation (No. XHD24B0301). Y.L. thanks Tong Jiang, and Liangdong Hu for helpful discussions about TD-DMRG and DDMRG. 
\end{acknowledgments}

% \nocite{*}

\bibliography{apssamp}% Produces the bibliography via BibTeX.

\end{document}